\documentclass[prb,twocolumn,superscriptaddress,showpacs,amsmath,amssymb]{revtex4}
\usepackage{amsfonts}
\usepackage{bbm}
\usepackage{graphicx}
\usepackage{dcolumn}
\usepackage{bm}

\begin{document}

\title{Magnetoanisotropic spin-triplet Andreev reflection in ferromagnet-Ising superconductor junctions}

\author{Peng Lv}
\affiliation{International Center for Quantum Materials, School of Physics, Peking University, Beijing 100871,
China}
\author{Yan-Feng Zhou}
\affiliation{International Center for Quantum Materials, School of Physics, Peking University, Beijing 100871,
China}
\author{Ning-Xuan Yang}
\affiliation{International Center for Quantum Materials, School of Physics, Peking University, Beijing 100871,
China}
\author{Qing-Feng Sun}
\email[]{sunqf@pku.edu.cn}
\affiliation{International Center for Quantum Materials, School of Physics, Peking University, Beijing 100871, China}
\affiliation{Collaborative Innovation Center of Quantum Matter, Beijing, 100871, China}
\affiliation{CAS Center for Excellence in Topological Quantum Computation, University of Chinese Academy of Sciences, Beijing 100190, China}

\begin{abstract}
We theoretically study the electronic transport through a ferromagnet-Ising superconductor junction.
A tight-binding Hamiltonian describing the Ising superconductor is presented.
Then by combing the non-equilibrium Green's function method,
the expressions of Andreev reflection coefficient and conductance are obtained.
A strong magnetoanisotropic spin-triplet Andreev reflection is shown, and
the magnetoanisotropic period is $\pi$ instead of $2\pi$ as in the conventional
magnetoanisotropic system.
We demonstrate a significant increase of the spin-triplet Andreev reflection
for the single-band Ising superconductor.
Furthermore, the dependence of the Andreev reflection on the incident energy and incident angle
are also investigated. A complete Andreev reflection can occur when the incident energy is equal
to the superconductor gap, regardless of the Fermi energy (spin polarization) of the ferromagnet.
For the suitable oblique incidence, the spin-triplet Andreev reflection can be strongly enhanced.
In addition, the conductance spectroscopies of both zero bias and finite bias are studied,
and the influence of gate voltage, exchange energy, and
spin-orbit coupling on the conductance spectroscopy are discussed in detail.
The conductance reveals a strong magnetoanisotropy with period $\pi$ as the Andreev reflection coefficient. When the magnetization direction is parallel to the junction plane,
a large conductance peak always emerges at the superconductor gap.
This work offers a comprehensive and systematic study of the spin-triplet
Andreev reflection, and has underlying application of $\pi$-periodic spin valve in spintronics.
\end{abstract}
\maketitle

\section{Introduction}

Ferromagnetism and superconductivity are mutually exclusive types of phases,
but the interplay between them can lead to a variety of interesting phenomena.\cite{A1,A2,A3}
The development in nanofabrication and material growth technologies makes it possible
to fabricate various kinds of hybrid mesoscopic structures, and the ferromagnet-superconductor
junctions have received widespread attention over the past few decades,\cite{B1,B2,B3,B4,B5}
both for its fundamental physics and potential applications.
Andreev reflection occurs in the conductor-superconductor interface,
where the incident electron from the conductor is reflected back as a hole
and a Cooper-pair is injected into the superconductor.\cite{AndR1,AndR2,AndR3,AndR4,AndR5,AndR6}
When the applied bias is less than the superconducting gap, the conductance of the
conductor-superconductor junction is mainly determined by the Andreev reflection.\cite{AndR2,AndR3}
For the ferromagnet-superconductor junction, it is revealed that the Andreev reflection is strongly affected by the spin polarization of the ferromagnet lead.\cite{B1}
In particular, when the ferromagnet is completely spin-polarized, the Andreev reflection
vanishes and the conductance becomes zero when the applied voltage is smaller than the superconducting gap.
This is because the Cooper pairs are formed by electrons from different spin sub-bands,
which usually have different density of states near the Fermi surface at the ferromagnet.
Especially, when the ferromagnet becomes completely spin-polarized,
the Andreev reflection is completely suppressed due to the presence of only one
spin sub-band near the Fermi surface.
Thus, it is possible to extract the information about spin polarization of ferromagnets
from the differential Andreev conductance,
which have been verified in many experiments.\cite{C1,C2}

However, there are some experiments indicating nonzero Andreev reflection even
when the ferromagnet is completely spin-polarized.\cite{D1,D2}
This can be attributed to the inhomogeneous magnetization or spin-flip mechanism of
the junction.\cite{E1,E2,E3} Recently, many efforts have been devoted to study
the Andreev reflection in ferromagnet-superconductor junction with interfacial
Rashba spin-orbit coupling.\cite{F1,F2,F3,F4} In hybrid systems with distinct electronic
structures and breaking space inversion symmetry, Rashba spin-orbital field can
be induced at the interface of the junction.\cite{G1,G2,G3}
In the presence of the interfacial spin-orbital field, spin-triplet Andreev reflection
occurs between the ferromagnet and the superconductor, where the incident electrons and
the reflected holes come from the same spin sub-band.
This spin-triplet Andreev reflection is strongly dependent on the direction of the
ferromagnet polarization. Some authors have proposed to use the Andreev reflection
spectroscopy for probing of the interfacial spin-orbital field,
with the effects of both s- and d-wave pairings being considered.\cite{F3,F4}

Recently, another kind of Cooper pairing, called Ising pairing, has been found in the experiments.\cite{IS1,IS2,IS3,IS4} This unconventional pairing arises from strong
intrinsic spin-orbit interaction and non-centrosymmetry in two-dimensional transition metal
dichalcogenides (TMDs), which leads to an effective Zeeman field that points
in the opposite directions at different valleys.
The observation of a large in-plane critical field and the enhancement of the Pauli paramagnetic
limit originate from the inter-valley pairing protected by spin-momentum locking.\cite{IS2,IS3}
The Ising spin-orbit coupling can also induce spin-triplet pairing in the in-plane directions,
which may be detected by the conductance spectroscopy of a half-metal/Ising superconductor
junction.\cite{H1,H2,H3}
More recently, some theoretical works indicate a rich phase diagram of unconventional
superconductivity in bilayer TMDs,\cite{I1,I2} and the effect of disorder
on the upper critical field have also being investigated.\cite{I3}

In this paper, by combing the tight-binding model with the non-equilibrium Green's function method,
we investigate the Andreev reflection in ferromagnet-Ising superconductor junctions.
A recent work has devoted to the charge conductance of the half-metal/Ising
superconductor junctions. They mention that the Andreev reflection can occur in one spin sub-band
due to the triplet pairing correlations.\cite{H3}
Here we systematically study the quantum transport through
the ferromagnet-Ising superconductor junctions.
The results show that the spin-triplet Andreev reflection is strong magnetoanisotropy
and the magnetoanisotropic period is $\pi$ instead of $2\pi$ as in the conventional
magnetoanisotropic system.
We demonstrate a significant increase of the spin-triplet Andreev reflection
when the Fermi energy is located between the spin-up and spin-down sub-bands
in the normal phase of the Ising superconductor.
When the incident energy is equal to the superconductor gap,
a complete Andreev reflection can occur regardless of the spin polarization $P$ of the ferromagnet.
For the suitable oblique incident case, the spin-triplet Andreev reflection can be strongly enhanced.
The conductance spectroscopies of both zero bias and finite bias are studied in detail,
and the influence of gate voltage, exchange energy, and spin-orbit coupling on the conductance spectroscopy are also discussed.
A strong magnetoanisotropy with period $\pi$ is shown in the conductance spectroscopy,
and a large conductance peak always emerges at the superconductor gap when
the magnetization direction is parallel to the plane of the ferromagnet-Ising superconductor junction.

The rest of this paper is organized as follows.
We will start in Sec. \ref{sec:Model} by demonstrating the Hamiltonian of
the ferromagnet-Ising superconductor junctions and showing the corresponding band structures.
In Sec. \ref{sec:transmission}, we calculate the Andreev reflection coefficient,
studying its dependence on the incident energy, Fermi energy of the Ising superconductor, the
transverse wave vector (the oblique incident case), and show the magnetoanisotropic properties.
In Sec. \ref{sec:conductance}, the conductance spectra are presented and
the influences of both exchange energy and spin-orbit coupling are investigated.
Finally, a brief summary is presented in Sec. \ref{sec:conclusions}.

\begin{figure}[!htb]
\includegraphics[width=1.0\columnwidth]{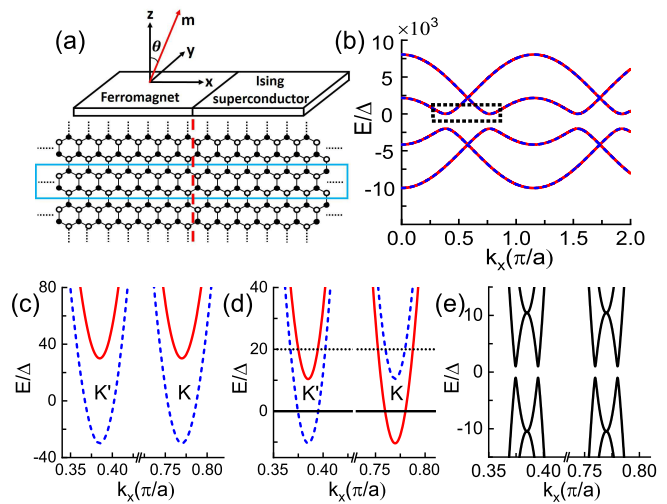}
\caption{ (Color online) (a) Schematic illustration and lattice diagram of
the ferromagnet-Ising superconductor junction.
$\theta$ is the angle between the magnetization vector $\bm{m}$ and the $z$ axis.
The system is assumed periodic along the $y$ direction and the blue box shows the unit cell. 
The nearest-neighbor distance of the hexagonal lattice is set to be $a$.
(b) Energy bands of the TMDs at the normal phase case while $k_{y}=0$.
Red lines denote the energy bands of spin-up electrons,
while the dashed blue lines are the spin-down bands.
(c) Energy bands of the ferromagnet near the Fermi surface $E_{F}=0$, in which
the spin degeneracy of the spin-up and spin-down bands is removed.
(d) Zoom-in figure of the energy bands of the dotted box in (b).
The black solid and dot lines schematically show the positions of Fermi energy
for the single-band and double-bands Ising superconductors, respectively.
(e) Energy bands of the Ising superconducting phase.
The relevant parameters are $\Delta= 1 meV$, $\epsilon=\beta=-1 eV$, $t=-3 eV$, $\beta_{s}=-2 meV$, $m=30 meV$,
and $E_{L}=E_{R}=0$.}
\label{fig1}
\end{figure}

\section{\label{sec:Model}Model and formalism}

In this section, the model Hamiltonian and the corresponding band structures are demonstrated.
We consider the system consisting of a ferromagnet coupled to a Ising superconductor lead
[as shown in Fig.\ref{fig1}(a)] with the Hamiltonian
\begin{align}\label{eq:a}
H=H_{IS}+H_{F}+H_{C},
\end{align}
where $H_{IS}$, $H_{F}$ and $H_{C}$ are the Hamiltonians of the Ising superconductor,
the ferromagnet, and the coupling between them, respectively.
When the Ising superconductor is in the normal phase, the broken sublattice symmetry and spin-orbit coupling give rise to an opposite spin sub-band splits at the K and K' valleys. At the K valley, spin-up band has lower energy than the spin-down band, but it is the opposite at the K' valley, that is, the spin-down band is lower [as shown in Fig.\ref{fig1}(d)].\cite{IS1,IS2,IS3,H3,AF1}
Therefore, we describe the Ising superconductor lead by using the following
hexagonal lattice tight-binding Hamiltonian $H_{IS}$
\begin{align}\label{eq:b}
&H_{IS}=\sum_{\bm{i}\sigma}(\epsilon-E_{R}+\lambda_{\bm{i}}\beta) b^{\dagger}_{\bm{i}\sigma} b_{\bm{i}\sigma}+ \sum_{\langle{\bm{i}\bm{j}}\rangle\sigma} t  b^{\dagger}_{\bm{i}\sigma}b_{\bm{j}\sigma} \nonumber \\
&+\sum_{\langle\langle{\bm{i}\bm{j}}\rangle\rangle \sigma \sigma^{'}}i\beta_{s}\nu_{\bm{i}\bm{j}}s_{\sigma \sigma^{'}}^{z}b_{\bm{i}\sigma}^{\dagger}b_{\bm{j}\sigma^{'}}
+\sum_{\bm{i}}\Delta b^{\dagger}_{\bm{i}\uparrow} b^{\dagger}_{\bm{i}\downarrow} +H.c.,
\end{align}
where $b_{\bm{i}\sigma}$ and $b^{\dagger}_{\bm{i}\sigma}$ are the annihilation
and creation operators at the discrete site $\bm{i}=(i_x,i_y)$ of the Ising superconductor lead,
with $\sigma=\uparrow, \downarrow$ denoting the electron's spin.
$\epsilon -E_R$ is the on-site energy, which can be modulated through
 the gate voltage in experiments. $\beta$ represents the energy difference
between A-B sublattice with $\lambda_{\bm{i}}=\pm1$ for A (B) sublattice.
The second term is the nearest-neighbor hopping term, and $t$ is the hopping energy.
The third term is the spin-orbit interaction which connects
second nearest neighbor. The same term also appears in the seminal work of the
quantum spin Hall effect in graphene.\cite{J1,J2,J3,AF2}
$s_{z}$ is a Pauli matrix representing the electron's spin and $\nu_{\bm{i}\bm{j}}=-\nu_{\bm{j}\bm{i}}=\pm1$ depending on the orientation of the two
site $\bm{i}$ to $\bm{j}$.
The spin-orbit coupling strength $\beta_{s}$ is usually very small comparing to $t$.
The fourth term in Eq.(\ref{eq:b}) describes the superconductivity and $\Delta$
is the superconducting gap.
Recent experiments show that the transition temperature in superconducting TMDs is
about $2\sim10 K$.\cite{IS1,IS2,IS3} Hence, it's reasonable to set $\Delta=1\ \rm meV$ as
the energy unit in the numerical calculation.

For the ferromagnet lead, a magnetization vector $\bm{m}$ is considered and
the Hamiltonian $H_{F}$ can be written as
\begin{align}\label{eq:c}
H_{F}=&\sum_{\bm{i}\sigma}(\epsilon-E_{L}+\lambda_{\bm{i}}\beta) a^{\dagger}_{\bm{i}\sigma} a_{\bm{i}\sigma}+ \sum_{\langle{\bm{i}\bm{j}}\rangle\sigma} t a^{\dagger}_{\bm{i}\sigma}a_{\bm{j}\sigma} \nonumber \\
&+\sum_{\bm{i} \sigma \sigma^{'}} \left( \bm{m} \cdot \bm{\sigma} \right)_{\sigma \sigma^{'}}  a_{\bm{i}\sigma}^{\dagger} a_{\bm{i}\sigma^{'}},
\end{align}
where $a_{\bm{i}\sigma}$ and $a^{\dagger}_{\bm{i}\sigma}$ are the annihilation
and creation operators at the discrete site $\bm{i}$ of the ferromagnet lead.
The first two terms are almost the same as those in $H_{IS}$ [see Eq.(\ref{eq:b})]
except for $E_{L}$, which can be adjusted independently through the gate voltage in the left side.
Here we have assumed that the ferromagnet lead is a hexagonal lattice model
as the Ising superconductor. The results are similar for a square lattice model.
$\bm{m}=m(\sin\theta, 0, \cos\theta)$ in the third term is the magnetization vector
in the ferromagnet lead, where $m$ is the exchange energy and $\theta$ is the angle
between $\bm{m}$ and the $z$ axis. The orientation of $\bm{m}$ can be continuously
changed by applying a small magnetic field,\cite{J4}
while the Ising superconductor remains unchanged due to the Meissner effect.

The Hamiltonian $H_{c}$ of the coupling between the ferromagnet lead and Ising superconductor is
\begin{align}\label{eq:d}
H_{T}=\sum_{\bm{i}\bm{j}\sigma}t_{c}a_{\bm{i}\sigma}^{\dagger}b_{\bm{j}\sigma}+h.c.,
\end{align}
where $t_c$ is the coupling strength. Here, we assume that the
couplings are only between the outermost layers of the left and right leads
[as shown in Fig.\ref{fig1}(a)], and set $t_{c}=t$ for convenience of calculations.

Here we consider that the system width in the $y$ direction is very large and the periodic boundary condition is adopted.
By applying the Bloch theorem along $y$-direction and introducing the corresponding wave vector $k_{y}$,
the Hamiltonian in Eq.(\ref{eq:a}) can be rewritten as $H=\sum_{k_{y}}H(k_{y})$,
where $H(k_{y})$ is the one-dimensional Hamiltonian for a given $k_{y}$.
Thus, by calculating the transmission coefficients of each $H(k_{y})$ and summing them up,
we can construct the transport properties of the whole two-dimensional system.
Here, we choose each unit cell containing four atoms in $y$-direction as illustrated
in the blue box in Fig.\ref{fig1}(a),
hence the Brillouin zone is one half smaller than the usual Brillouin zone of
graphene.\cite{Graph1,Graph2,Graph3}
In Fig.\ref{fig1}(b), we plot the energy bands of the TMDs in the normal phase
[corresponding to $H_{IS}$ in Eq.(\ref{eq:b}) with $\Delta=0$] at $k_y=0$.
We mainly consider the electrons transport near the Fermi surface,
and the four lower energy bands can be ignored, as they are much below the Fermi energy $E_{F}=0$.
In Fig.\ref{fig1}(d), the zoom-in figure of the energy bands near the K and K' points in Fig.\ref{fig1}(b)
are presented. The spin-orbit coupling can be viewed as an effective magnetic field $\bm{B}$
that points in the opposite directions at the K and K' points.
At K point, spin up electrons have lower energy while at K' point the other way around.
These energy bands are agreement with the Ising superconductor in the normal case,\cite{IS1,H3}
which means that the tight-binding Hamiltonian $H_{IS}$ in Eq.(\ref{eq:b}) does well describe
the Ising superconductor.
Experiments indicate that the energy interval between spin-up and spin-down sub-bands is
about $10\sim20\ \rm meV$,\cite{IS1,IS2,IS3} and we set this energy interval to be $20\Delta$
in the calculation below [see Fig.\ref{fig1}(d)].
On the other hand, the energy bands of the ferromagnet lead are displayed in Fig.\ref{fig1}(c).
The spin degeneracy of the spin-up and spin-down sub-bands is lifted,
and the magnetization vector $\bm{m}$ induces a Zeeman-type field that points in the same
direction at both K and K' valleys.
Note that in Fig.\ref{fig1}(c) the Fermi energy $E_{F}=0$ is located in the middle of the spin-up
and spin-down sub-bands. Owing to the absence of the spin-up sub-band near the Fermi surface,
the spin polarization $P=|\frac{\rho_{\uparrow}-\rho_{\downarrow}}{\rho_{\uparrow}+\rho_{\downarrow}}|$
is $1$ at $E_{F}$, where $\rho_{\uparrow(\downarrow)}$ is the
density of states of spin-up (down) electrons.
In this case, the ferromagnet is completely spin-polarized and
the ordinary Andreev reflection is completely suppressed.
Once the Fermi energy $E_{F}$ is shifted inside the spin-up sub-bands,
the polarization $P$ quickly drops to zero.
This is because the density of states for parabolic dispersion relation
in two dimension is a constant, irrespective of the Fermi wavelength.
Fig.\ref{fig1}(e) demonstrates the energy bands of the TMDs in the superconducting phase,
where the Ising pairing is created between the opposite spin sub-bands at K and K' valleys.
When the Fermi surface is located in the middle of the spin-up and spin-down
sub-bands of K (K') valley, only a single band at each valley participates in the paring process.
In general, at each valley, both two spin sub-bands will participate in the paring process.
Below, we will study both double-bands and single-band Ising superconductivities.
Experimentally, by tuning the Fermi surface of the TMDs through the gate voltage, the Ising superconductor can be either double-bands or single-band.
In addition, even if for a double-bands Ising superconductor,
as $k_{y}$ increases, the corresponding energy bands of $H(k_{y})$ move up,
and for some certain range of $k_{y}$, the Fermi energy $E_{F}$ enters into the energy
interval between the spin-up and spin-down sub-bands, which is similar to the single-band Ising superconductivity. With these considerations in mind, we systematically investigate
the electronic transport through a ferromagnet-Ising superconductor junction below.

Due to the wave vector $k_{y}$ being a good quantum number,
$H(k_{y})$ can be regarded as the Hamiltonian of a 1D ferromagnet-Ising superconductor nanowire
for a given $k_{y}$. The current flowing through the nanowire can be calculated from the evolution of total
number operator for the electrons in the left ferromagnet lead.\cite{B3,Asun1,Asun2,J5}
\begin{align}\label{eq:e}
I({k_y})&=-e\langle \frac{d}{dt}\sum_{i_x\in F,\sigma} a_{i_x, \sigma}^{\dagger}(k_y)a_{i_x, \sigma}(k_y) \rangle   \nonumber \\
&=\frac{2e}{\hbar} {\rm Re}{\sum_{i_x\in C,j_x\in F}} \mathrm{Tr} \left[G_{i_x, j_x}^{<}t_{j_x,i_x} \right]_{ee},
\end{align}
where $a_{i_x, \sigma}^{\dagger}(k_y)$ and $a_{i_x, \sigma}(k_y)$ are the Flourier
transformation of  $a_{\bm{i}, \sigma}^{\dagger}$ and $a_{\bm{i}, \sigma}$ along $y$-direction.
Here, $i_x\in C$ and $j_x\in F$ represent the site index that belong to the center region
and left ferromagnet lead respectively.
The center region can be chosen arbitrarily, which will not affect the final result.
Note that we have expressed various kinds of Green's functions in a generalized $4\times4$ Nambu representation, and $\mathrm{Tr} \left[\right]_{ee}$
is the trace acting on the particle space. By applying the Dyson equation, $G_{i_x, j_x}^{<}$ can be decoupled into the product of the unperturbed Green's function
of the ferromagnet and the Green's function of the center region.
Therefore, the current expression can be rewritten as
\begin{align}\label{eq:f}
I({k_y})=-\frac{2e}{\hbar}{\rm Im}\int\frac{d\omega}{2\pi}\mathrm{Tr} \left[\bm{G^{r}} \bm{f_{L}} \bm{\Gamma^{L}} +\frac{1}{2}\bm{G^{<}}\bm{\Gamma^{L}} \right]_{ee},
\end{align}
where $\bm{G^{r(<)}}$ is the retarded (lesser) Green's functions in the center region and $\bm{f_{L}}\equiv diag(f_{Le},f_{Le},f_{Lh},f_{Lh})$. $f_{Le}=f(\omega-eV)$
and $f_{Lh}=f(\omega+eV)$ are the Fermi distribution functions of particles and holes
of the left ferromagnet lead, with $V$ being the bias voltage.
We also introduce the linewidth matrices in the generalized Nambu representation $\bm{\Gamma^{L}}(\omega)=i[\bm{\Sigma}_{L}^{r}(\omega)-\bm{\Sigma}_{L}^{a}(\omega)]$,
where $\bm{\Sigma}_{L}^{r/a}(\omega)$ is the self energy due to the left ferromagnet lead.
By utilizing the Keldysh equation $\bm{G}^{<}=\bm{G}^{r}\bm{\Sigma}^{<}\bm{G}^{a}$,
$\bm{G}^{r}-\bm{G}^{a}=\bm{G}^{r}(\bm{\Sigma}^{r}-\bm{\Sigma}^{a})\bm{G}^{a}$,
and the self-energies $\bm{\Sigma}^{r,a,<}=\bm{\Sigma}_{L}^{r,a,<}+\bm{\Sigma}_{R}^{r,a,<}$,
the tunneling current $I(k_y)$ is reduced to\cite{Asun1,Asun2}
\begin{align}\label{eq:g}
I({k_y})=I_{N}({k_y})+I_{A}({k_y}),
\end{align}
with
\begin{eqnarray}
I_{N}({k_y})&=&\frac{e}{h}\int d\omega \left[f_{Le}-f_{R}\right]
\mathrm{Tr} \left[\bm{\Gamma}_{ee}^{L}\left(\bm{G}^{r}\bm{\Gamma}^{R}\bm{G}^{a}\right)_{ee}\right],\\
I_{A}({k_y})&=&\frac{e}{h}\int d\omega \left[f_{Le}-f_{Lh}\right] \mathrm{Tr} \left[\bm{G}_{eh}^{r} \bm{\Gamma}_{hh}^{L} \bm{G}_{he}^{a} \bm{\Gamma}_{ee}^{L}\right],
\end{eqnarray}
where $f_{R}=f(\omega)$ is the Fermi distribution function of the right Ising superconductor lead.
$\mathrm{Tr} \left[\bm{\Gamma}_{ee}^{L}\left(\bm{G}^{r}\bm{\Gamma}^{R}\bm{G}^{a}\right)_{ee}\right]\equiv T_{N}(k_y, \omega)$ is the normal tunneling coefficient caused by the quasiparticle transport, and $\mathrm{Tr} \left[\bm{G}_{eh}^{r} \bm{\Gamma}_{hh}^{L} \bm{G}_{he}^{a} \bm{\Gamma}_{ee}^{L}\right]\equiv T_{A}(k_y, \omega)$ is the Andreev reflection coefficient.\cite{Asun1,Asun2}
The total current can be obtained by summing up the contribution of each $I({k_y})$ and we finally get $I_{tot}=\sum_{k_y}I(k_y)$.
Then, the linear conductance is given by $G=\mathrm{lim}_{V\rightarrow 0}dI_{tot}/dV$.

\begin{figure}[!htb]
\includegraphics[width=1.0\columnwidth]{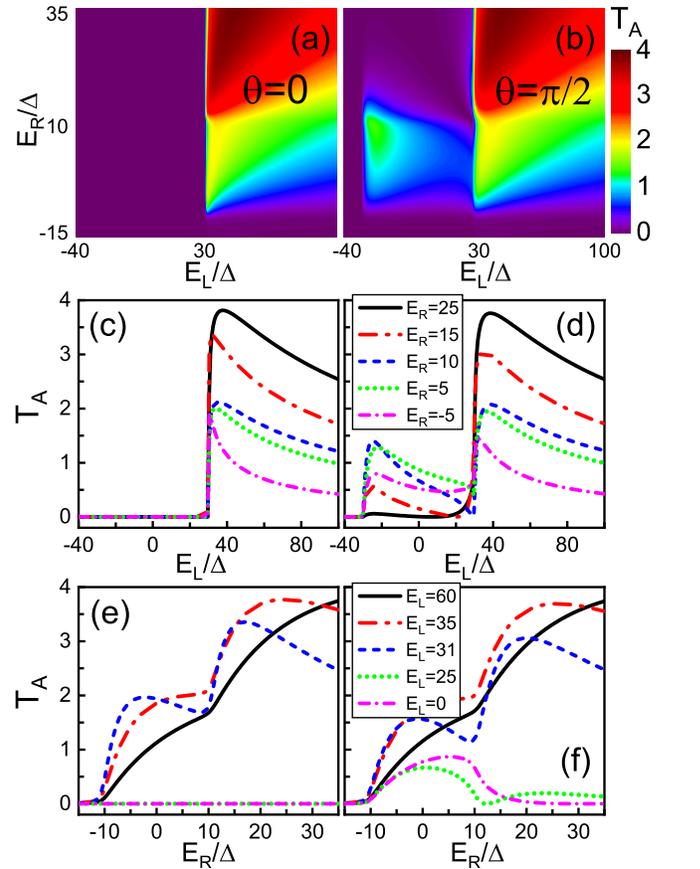}
\caption{(Color online) (a) and (b) are the 2D plot of Andreev reflection coefficient $T_{A}$
versus the on-site energy $E_{L}$ and $E_{R}$, with the incident energy $\omega=0$, $k_y=0$,
$\theta=0$ in (a) and $\theta=\pi/2$ in (b).
(c) and (d) are truncated intersector curves of $T_{A}$ versus $E_{L}$ in (a) and (b)
with $E_{R}=25$, $15$, $10$, $5$, and $-5$ respectively.
(e) and (f) are truncated intersector curves of $T_{A}$ versus $E_{R}$ in (a) and (b)
with $E_{L}=60$, $35$, $31$, $25$, and $0$ respectively.
Other parameters are the same as those in Fig.\ref{fig1}.}
\label{fig2}
\end{figure}

\section{\label{sec:transmission} Magnetoanisotropic spin-triplet Andreev reflection}

In this section, we investigate the Andreev reflection of the ferromagnetic-Ising
superconductor junction in detail.
Fig.\ref{fig2}(a) and (b) display the 2D plot of the Andreev reflection coefficient
$T_{A}(k_y=0, \omega=0)$ as a function of the on-site energy $E_{L}$ and $E_{R}$.
In Fig.\ref{fig2}(a) where the magnetization vector $\bm{m}$ is in the out-of-plane
directions with $\theta=0$,
once the ferromagnet is completely spin-polarized when $E_{L}$ is less than $m=30$,
$T_{A}$ quickly drops to zero which means that the ordinary Andreev reflection is completely suppressed.
However, when $\bm{m}$ is parallel to the in-plane directions with $\theta=\pi/2$,
$T_{A}$ survives even in the complete spin-polarized region [see Fig.\ref{fig2}(b)],
owing to the presence of the spin-triplet Andreev reflection.
This novel Andreev reflection can be understood as follow.\cite{J6}
Incident electron with its spin pointing to in-plane directions splits up into
two coherent electronic states in spin-up and spin-down channels via the ferromagnet-Ising
superconductor interface. The two electronic states
transform into two coherent hole states moving backward due to the Andreev reflection.
The coherent hole states with spin-up and spin-down project
back into the hole band of the left ferromagnet lead,
giving rise to the novel Andreev reflection. In this spin-triplet Andreev reflection,
both the incident electron and backward hole are in the same spin sub-band.

To further investigate the characteristics of the Andreev reflection in the ferromagnet-Ising
superconductor junction, we plot the truncated intersector curves of the Andreev
reflection coefficient $T_A$
as a function of the on-site energy $E_{L}$ in Fig.\ref{fig2}(c) and (d).
In Fig.\ref{fig2}(c) where $\theta=0$ (magnetization vector ${\bf m}$ being
perpendicular to the plane), $T_A$ is exactly zero when $E_L <|m|=30$ with
the ferromagnet lead being complete spin-polarized. In this region, the Andreev reflection
completely disappears.
$T_A$ quickly rises when $E_L$ passes $|m|$, reaches its maximum near $E_L=40$,
and gradually decreases for larger $E_{L}$ due to the Fermi wavelength mismatch.
Because of the double valleys and two spin degrees of freedom near the Fermi surface,
the in gap Andreev reflection coefficient is 4 for a perfect junction.
Here, $T_A$ is almost perfect for $E_{R}=25$, and decreases as $E_{R}$ decreases.
In particular, at the transition point of the double-bands and single-band Ising superconductivity
where $E_R=10$, $T_A$ approximately decreases by half.
This is because the alternative outgoing channels are just reduced by half from double-bands
to single-band Ising superconductivity.
While in Fig.\ref{fig2}(d) with $\theta=\pi/2$ (${\bf m}$ being parallel to the plane),
$T_A$ has a considerable value even for the completely spin-polarized left lead with $E_L <|m|$,
because that the spin-triplet Andreev reflection occurs.
The spin-triplet Andreev reflection is rather small for the double-bands Ising superconductor [e.g. see the curve of $E_R=25$ in Fig.\ref{fig2}(d)], and is enhanced for smaller $E_R$.
But for the single-band Ising superconductor, the spin-triplet Andreev reflection is
dramatically enhanced [see the curves of $E_R\leq 10$ in Fig.\ref{fig2}(d)].
To better illustrate this, we also plot the truncated intersector curves of $T_A$
as a function of $E_{R}$ in Fig.\ref{fig2}(e) and \ref{fig2}(f).
For $\theta=0$ in Fig.\ref{fig2}(e),  $T_A$ clearly exhibits a step behavior at $E_R=10$, and vanishes for the completely spin-polarized ferromagnet (see the curves of $E_L<|m|=30$).
However, with $\theta=\pi/2$ in Fig.\ref{fig2}(f),
forming of the single-band Ising-superconductivity reduces the ordinary Andreev reflection by half,
but meanwhile dramatically enhances the spin-triplet Andreev reflection.
As pointed out by previous works,\cite{H1,H2,H3} the Ising superconductivity induces a
spin-triplet pairing characterized by correlation function $\bm{d}=(0, 0, d_z)$, where
$|d_z(\omega)|=|\frac{2\xi\tilde{\beta}}
{[\omega^2-(\xi+\tilde{\beta})^2-\Delta^2][\omega^2-(\xi-\tilde{\beta})^2-\Delta^2]}|$.
In this expression, $\xi=\frac{p^2}{2m}-E_F$ is the kinetic energy measured
from the Fermi surface $E_F$ and $2\tilde{\beta}$ is the energy interval between spin-up and
spin-down sub-bands, which is about $20$ for the parameters in Fig.\ref{fig1}.
$|d_z|$ corresponds to the spin-triplet pairing with zero total spin projection,
and it's because of the nonzero $|d_z|$ that the novel spin-triplet Andreev reflection can occur.
Detailed analysis of $|d_z|$ with $\omega=0$ reveals that $|d_z|$ is maximum at
$\xi=-\tilde{\beta}$ where the single-band Ising superconductivity appears,
and cubic decays for large $\xi$.
Therefore, for double-bands Ising superconductivity with $E_R> \tilde{\beta} \approx 10$,
the spin-triplet Andreev reflection is rather weak,
but dramatically enhanced once $E_R< \tilde{\beta} \approx 10$
where the single-band Ising superconductivity is formed.

\begin{figure}[!htb]
\includegraphics[width=1.0\columnwidth]{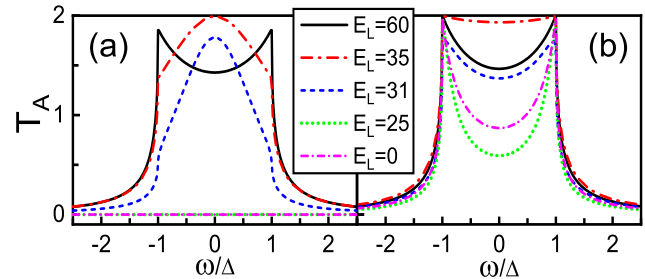}
\caption{(Color online) The Andreev reflection coefficient $T_{A}$ versus energy $\omega$
of incident electron for different energy $E_{L}$, with $\theta=0$ in (a) and $\theta=\pi/2$ in (b).
The parameters are $k_{y}=0$, $E_{R}=5$, and others are the same as those in Fig.\ref{fig1}.}
\label{fig3}
\end{figure}

To better appreciate the spin-triplet Andreev reflection,
we focus on the single-band Ising superconductivity and
present the Andreev reflection coefficient $T_{A}$ versus
the energy $\omega$ of the incident electron, as shown in Fig.\ref{fig3}.
In Fig.\ref{fig3}(a) with $\theta=0$ (${\bf m}$ being
perpendicular to the plane), $T_A$ undergoes a transition from a zero-bias dip to
a zero-bias peak as the energy $E_L$ decreases,
and is always zero for the completely spin-polarized ferromagnet with $E_L <|m|$
due to the absence of the spin-triplet Andreev reflection at $\theta=0$.
While in Fig.\ref{fig3}(b) with $\theta=\pi/2$ (${\bf m}$ being parallel to the plane),
$T_A$ always presents a zero-bias dip, irrespectively of the energy $E_L$ and
the spin polarization of the ferromagnet lead.
We emphasize that here the spin-triplet Andreev reflection has the same magnitude
as the ordinary Andreev reflection, especially near the gap edge where the incident energy $\omega=\Delta$.
In previous works, the in gap spin-triplet Andreev reflection is rather weak compared with
the ordinary Andreev reflection.
Here, we find that by tuning the on-site energy $E_R$ through the gate voltage
(which is equivalent to tune the Fermi surface),
the spin-triplet Andreev can be dramatically enhanced and reaches the maximum $2$ at $\omega=\Delta$.

\begin{figure}[!htb]
\includegraphics[width=1.0\columnwidth]{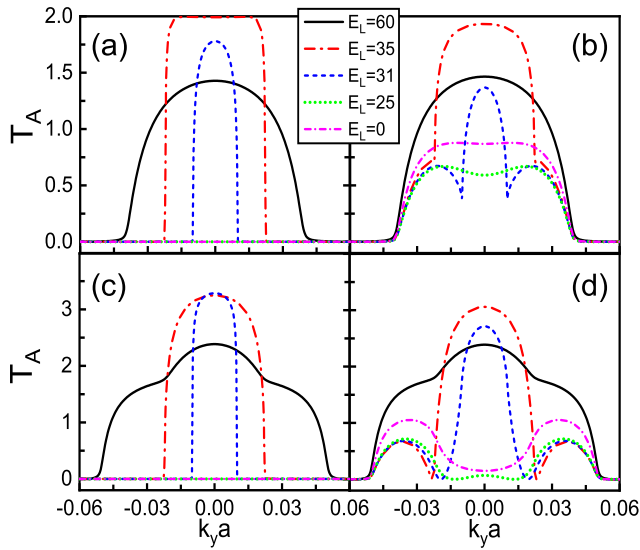}
\caption{ (Color online) The Andreev reflection coefficient $T_{A}$ versus transverse wave vector
$k_{y}$ for different energy $E_{L}$, with $\theta=0$ in [(a),(c)] and $\theta=\pi/2$ in [(b),(d)].
The parameters are taken to be $\omega=0$, $E_{R}=5$ in [(a),(b)] and $E_{R}=15$ in [(c),(d)].
Other parameters are the same as those in Fig.\ref{fig1}.}
\label{fig4}
\end{figure}

In the above, we study the normal incident case with the transverse wave vector $k_y =0$.
When $k_y\not= 0$, the oblique incidence occurs.
In Fig.\ref{fig4}, we present the Andreev reflection coefficient $T_{A}$ as a function of
the wave vector $k_{y}$ for the single-band Ising superconductor with $E_R=5$ [(a) and (b)] and
double-bands Ising superconductor with $E_{R}=15$ [(c) and (d)].
In Fig.\ref{fig4}[(a),(c)] with $\theta=0$, $T_A$ is always zero for the completely
spin-polarized ferromagnet with $E_L <|m|$, regardless of the values of $k_y$
(the normal and oblique incident cases) and $E_R$ (the double-bands and single-band superconductors),
because that the spin-triplet Andreev reflection can not occur at $\theta=0$.
While $E_L >|m|$, $T_A$ appears due to the ordinary Andreev reflection occurs.
$T_A$ has a large value at $k_y=0$. As $|k_{y}|$ increases, $T_A$ gradually drops,
and it quickly drops to zero once the spin polarization $P=1$ for a certain range of $k_{y}$.
On the other hand, as for $\theta=\pi/2$ in Fig.\ref{fig4}[(b),(d)],
$T_A$ appears even for the completely spin-polarized ferromagnet ($E_L<|m|$),
and it can also persist in large $k_{y}$ due to the presence of the spin-triplet Andreev reflection.
This indicates that the Andreev conductance can be enhanced by
changing the magnetization orientation of the ferromagnet.
In Fig.\ref{fig4}(b), the spin-triplet Andreev reflection is relatively large for various $k_y$,
due to the formation of the single-band Ising superconductivity.
While in Fig. \ref{fig4}(d), the spin-triplet Andreev reflection is rather small for $k_ya \in [-0.02, 0.02]$, but dramatically enhanced in the regions $k_ya \in [-0.05, -0.02]$ and $[0.02, 0.05]$.
This is because at critical point $k_ya=\pm0.02$, the Fermi surface starts to move into the energy
interval between the spin-up and spin-down sub-bands of the Ising superconductor.
This is different from the ordinary Andreev reflection where oblique incident modes generally
suppress the Andreev reflection coefficient.
Here, for the double-bands Ising superconductivity, the main contribution of
the spin-triplet Andreev reflection comes from the oblique incident modes [see Fig.\ref{fig4}(d)].

\begin{figure}[!htb]
\includegraphics[width=1.0\columnwidth]{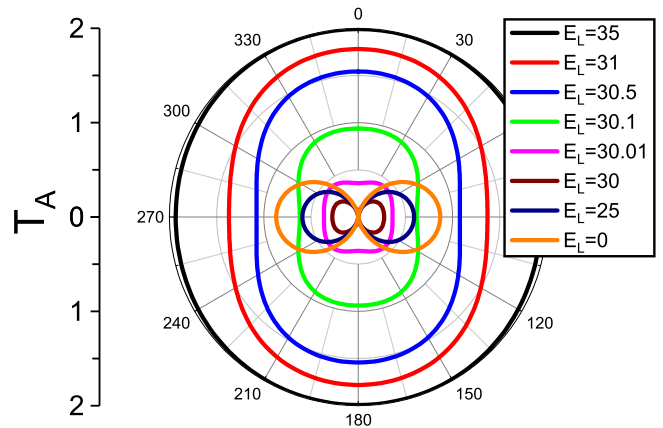}
\caption{
(Color online) Polar plot of the Andreev reflection coefficient $T_{A}$
as a function of the magnetization orientation $\theta$.
Different colours indicate different energy $E_{L}$.
The parameters are $E_R=5$, $k_y=0$, and $\omega=0$. Other parameters are the same as those in Fig.\ref{fig1}.}
\label{fig5}
\end{figure}

At the last of this section, we investigate the magnetoanisotropy of the spin-triplet Andreev reflection.
Fig.\ref{fig5} displays the Andreev reflection coefficient $T_A$ as a function of the
magnetization orientation $\theta$ for various energy $E_L$.
For $E_L$ much larger than $|m|$ (e.g. $E_L=35$)
where the spin polarization $P$ of the ferromagnet lead is about $0$,
the Andreev reflection is nearly isotropic and $T_A$ is nearly same for various magnetization directions.
As $E_L$ decreases with increasing polarization $P$ of the ferromagnet,
$T_A$ gradually reduces, meanwhile the magnetoanisotropy of $T_A$ becomes significant,
which exhibits its maximum at $\theta=0$ and minimum at $\theta=\pi/2$.
Near the completely spin-polarized transition point ($E_L=|m|=30$),
$T_A$ quickly shrinks, and manifests itself as a strong magnetoanisotropy
at the completely spin-polarized region,
where $T_A=0$ at $\theta=0$ and exhibits its maximum at $\theta=\pi/2$.
Note that $T_A$ is symmetrical about $\theta=\pi/2$ [$T_A(\frac{\pi}{2} +\theta)=T_A(\frac{\pi}{2} -\theta)$]
and shows a $\pi$-periodic oscillation behavior [$T_A(\theta+\pi)=T_A(\theta)$].
In fact, the spin-orbit coupling in the Ising superconductor can be viewed as a spin-flip mechanism.
At $\theta=0$, the spin in the $z$ direction is conserved in the scattering process,
and the lack of spin-flip results in the absence of the spin-triplet Andreev reflection.
For nonzero $\theta$, $s_z$ is not a conserved quantity anymore, and spin-flip mechanism begins
to take effect. With increasing $\theta$ from $0$, the spin-flip strength increases fist, reaches
its maximum at $\theta=\pi/2$, then decreases and finally vanishes at $\theta=\pi$,
where $s_z$ becomes a conserved quantity again.
So the change of $T_A$ with $\theta$ is not monotonic and exhibits a $\pi$-periodic oscillation.

\section{\label{sec:conductance} Magnetoanisotropic conductance and angular dependence}

\begin{figure}[!htb]
\includegraphics[width=1.0\columnwidth]{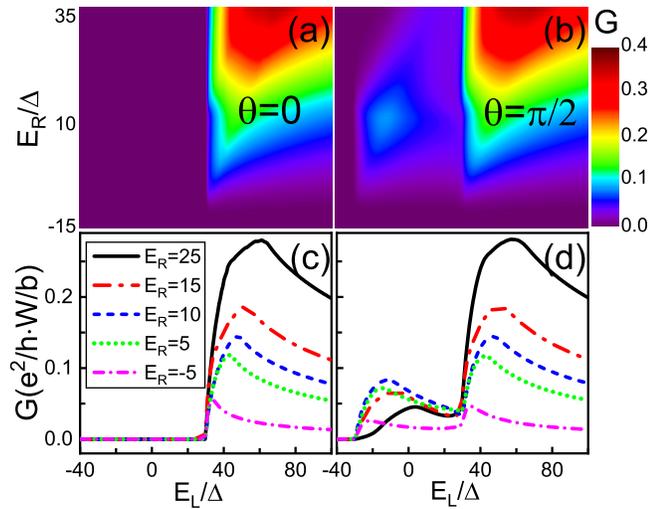}
\caption{
(Color online) (a) and (b) are the 2D plot of the linear conductance $G$ versus
the energy $E_{L}$ and $E_{R}$, with magnetization direction $\theta=0$ in (a) and $\theta=\pi/2$ in (b).
(c) and (d) are truncated intersector curves of $G$ versus $E_{L}$ in (a) and (b)
with $E_{R}=25$, $15$, $10$, $5$, and $-5$, respectively. Here, $W$ is the width of the ferromagnet-Ising
superconductor junction, and $b=3a$ is the width of the unit cell of the blue box in Fig.\ref{fig1}(a).
Other parameters are the same as those in Fig.\ref{fig1}.}
\label{fig6}
\end{figure}

Actual experiments are mainly concerned about the differential conductance
rather than the transmission coefficients.
So in this section, we investigate the conductance spectra of the ferromagnet-Ising superconductor junction.
Fig.\ref{fig6}(a) and \ref{fig6}(b) show a 2D plot of the linear conductance $G$ versus
the on-site energy $E_{L}$ and $E_{R}$.
In Fig.\ref{fig6}(a) where $\bm{m}$ is along the out-of-plane direction ($\theta=0$),
the linear conductance can reach $0.4$ when $E_L>|m|=30$ with the polarization $P\approx 0$,
but rapidly reduces to zero at the completely spin-polarized region ($E_L<|m|$)
due to the complete suppression of the Andreev reflection [also see Fig.\ref{fig6}(c)].
However, when $\bm{m}$ is along the in-plane directions ($\theta=\pi/2$)
as illustrated in Fig.\ref{fig6}(b), the spin-triplet Andreev reflection occurs and the conductance
is nonzero even at the completely spin-polarized region ($E_L<|m|$).
For the single-band Ising superconductor ($E_R<\tilde{\beta}\approx 10$), the conductance $G$
has the value about from $0.05$ to $0.1$. The conductance $G$ gradually increases with the increase of $E_R$,
and it reaches the maximum around $E_R=\tilde{\beta}$
(the transition point of the single-band and double-bands Ising superconductors).
Then with the further increase of $E_R$, $G$ gradually reduces again.
For clarity, we also present the truncated intersector curves of $G$ versus $E_{L}$ for various $E_R$,
as shown in Fig.\ref{fig6}(c) and (d).
When $E_L>|m|$ with the polarization $P\approx 0$,
the conductances are almost the same for $\theta=0$ and $\pi/2$.
However for $E_L <|m|$ with the completely spin-polarized ferromagnet,
the conductance is zero at $\theta=0$ due to the suppression of the Andreev reflection,
and it has a considerable value at $\theta=\pi/2$ because of
the occurrence of the spin-triplet Andreev reflection.
Thus, in order to clearly observe this novel conductance in real experiments,
it's beneficial to choose a highly spin-polarized ferromagnet lead.

\begin{figure}[!htb]
\includegraphics[width=1.0\columnwidth]{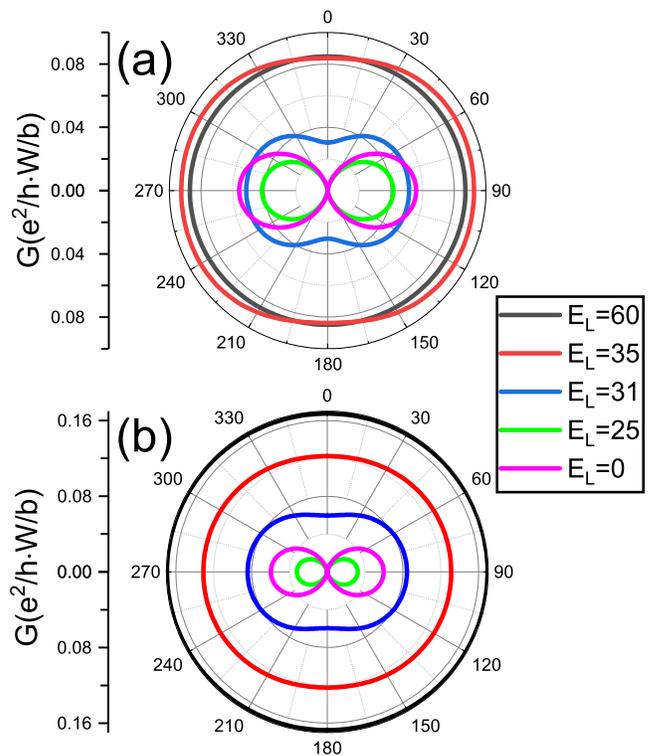}
\caption{
(Color online) Polar plot of the linear conductance $G$ as a function of the magnetization
orientation $\theta$, with the energy $E_R=5$ in (a) and $E_R=15$ in (b) corresponding to
the single-band and double-bands Ising superconductors.
Different colours indicate different energy $E_{L}$.
Other parameters are the same as those in Fig.\ref{fig1}.}
\label{fig7}
\end{figure}

Due to the magnetoanisotropy of the Andreev reflection coefficient,
the linear conductance shows similar magnetoanisotropic behaviors, as illustrated in Fig.\ref{fig7}.
In Fig.\ref{fig7}(a) where single-band Ising superconductivity is formed,
the linear conductance $G$ exhibits a small magnetoanisotropy for large $E_L$
with the spin polarization $P\approx 0$ (e.g. see $E_L=60$).
As $E_L$ decreases, the conductance $G$ reduces and the magnetoanisotropy becomes more significant.
When $E_L<|m|$ with the spin polarization $P=1$,
the conductance $G$ is highly magnetoanisotropic, where $G$ vanishes at $\theta=0$ and presents
its maximum at $\theta=\pi/2$, exhibiting a $\pi$-periodic oscillatory behavior.
On the other hand, in the case of double-bands Ising superconductivity as
demonstrated in Fig.\ref{fig7}(b), the magnetoanisotropy of the conductance is
relatively weaker than that in Fig.\ref{fig7}(a) when $E_L>|m|$,
but still reveals similar features at the completely spin-polarized region with $E_L<|m|$.
In this case, the oblique incident spin-triplet Andreev reflection dominates 
the conductance [see Fig.\ref{fig4}(d)].
Note that this oscillatory period is different from that in conventional tunnel magnetoresistance
(TMR),\cite{K1,K2,K3,AF3} where the conductance is maximum in the spin parallel state,
and drops to minimum in the spin anti-parallel state.
As the magnetization orientation changes, the TMR exhibits a $2\pi$-periodic oscillatory behavior.
Here, the conductance in the ferromagnet-Ising superconductor junction exhibits a $\pi$-periodic
oscillatory behavior, attributing to the spin-orbit coupling induced spin-triplet Andreev reflection
in the interface of the junction.

\begin{figure}[!htb]
\includegraphics[width=1.0\columnwidth]{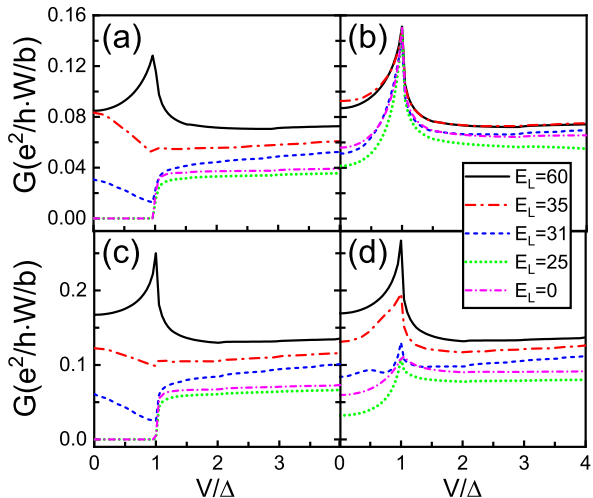}
\caption{
(Color online) The finite bias conductance spectra of the ferromagnet-Ising superconductor junction.
The energy $E_{R}$ is $5$ in [(a),(b)] and $15$ in [(c),(d)] corresponding to
the single-band and double-bands Ising superconductors, respectively.
The magnetization orientation $\theta=0$ in [(a),(c)] and $\theta=\pi/2$ in [(b),(d)].
Other parameters are the same as those in Fig. \ref{fig1}.}
\label{fig8}
\end{figure}

Let us study the finite bias conductance of the ferromagnet-Ising superconductor junction.
In Fig.\ref{fig8}[(a),(c)] where $\bm{m}$ is along the out-of-plane direction ($\theta=0$),
the conductance exhibits a peak at the gap edge $V=\Delta$ for $E_L$ much larger than $|m|$
(see the curve of $E_L=60$).
As $E_L$ decreases, the in gap conductance $G$ decreases and the peak at $V=\Delta$ gradually
transforms to a dip.
Finally when $E_L<|m|$ with the completely spin-polarized ferromagnet,
$G$ completely vanishes at the region of $V<\Delta$ due to the absence of the ordinary Andreev reflection.
However, in Fig.\ref{fig8}[(b),(d)] where $\bm{m}$
is along the in-plane directions ($\theta=\pi/2$),
the in gap conductance is dramatically enhanced, especially at the gap edge.
Even for the completely spin-polarized case ($E_L<|m|$), the conductance still has a considerable value
in the gap due to the occurrence of the spin-triplet Andreev reflection.
In Fig.\ref{fig8}(b) where single-band Ising superconductivity is formed,
the conductance at $V=\Delta$ pins at $G\approx 0.15$.
On the other hand, in the case of double-bands Ising superconductivity in Fig.\ref{fig8}(d),
the conductance peak gradually decreases as $E_L$ decreases, but the conductance peak always persists at $V=\Delta$, irrespective of the energy $E_L$.

\begin{figure}[!htb]
\includegraphics[width=1.0\columnwidth]{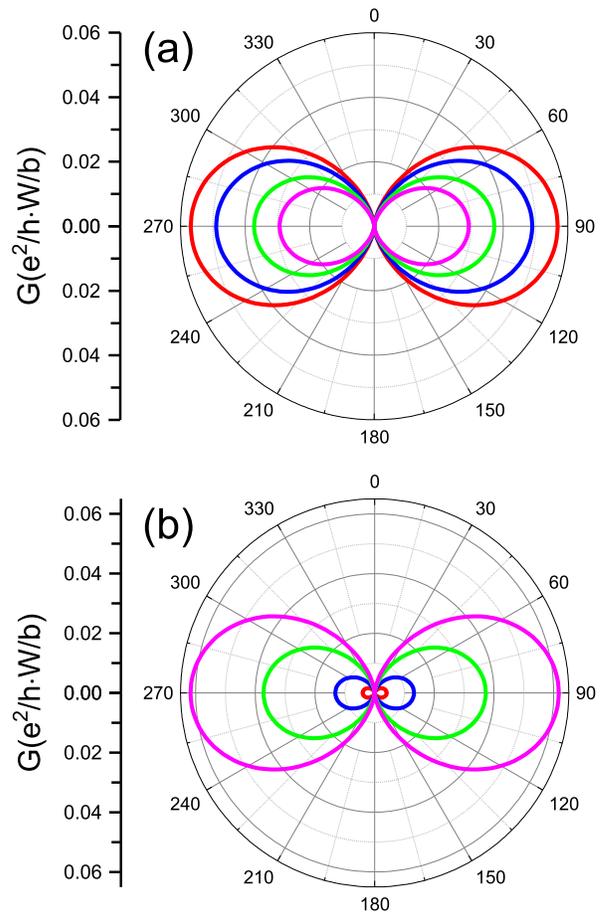}
\caption{
(Color online) Polar plot of the linear conductance $G$ as a function of the
magnetization orientation $\theta$.
(a) for different exchange energy with $m=10$, $20$, $30$, and $40$ from outmost to innermost.
(b) for different spin-orbit coupling strength with $\beta_{s}=-0.5$, $-1$, $-2$, and $-3$
from innermost to outmost. The parameters are the same as those in Fig.\ref{fig1}}.
\label{fig9}
\end{figure}

Finally, let us investigate the influence of exchange energy and spin-orbit coupling
on the linear conductance $G$.
From Fig.\ref{fig9}(a), we can see that the strong magnetoanisotropy with $\pi$-periodic
behavior can well survive regardless of the magnetization strength $m$.
$G$ is zero at $\theta=0$ and is maximum at $\theta=\pi/2$.
Increasing the magnetization strength results in a linearly decrease of the conductance $G$,
due to the Fermi wavelength mismatch at two sides.
In Fig.\ref{fig9}(b), $G$ versus $\theta$ for the different spin-orbit strength $\beta_s$ are shown.
The conductance $G$ still reveals the strong magnetoanisotropy for all $\beta_s$.
In addition, $G$ can be greatly enhanced by strengthening the spin-orbit coupling.
This is because that the increase of spin-orbit coupling enlarges the energy interval
of the spin-up and spin-down sub-bands of TMDs [see Fig.\ref{fig1}(d)].
So in real experiments, the spin-orbit interaction plays a important role in detecting
this novel spin-triplet Andreev reflection.

\section{\label{sec:conclusions}Summary}

In conclusion, we theoretically investigate the electronic transport through a
ferromagnet-Ising superconductor junction.
By combing the tight-binding model with the non-equilibrium Green's function method,
the expressions of Andreev reflection coefficient and conductance are obtained.
The results show that they reveal a strong magnetoanisotropy due to the spin-triplet Andreev reflection.
The magnetoanisotropic period is $\pi$ instead of $2\pi$ in the conventional
magnetoanisotropic system.
For the completely spin-polarized ferromagnet, the Andreev reflection disappears
when the magnetization direction is perpendicular to the junction plane, but it has a considerable
value when the magnetization direction is parallel to the junction plane due to the occurrence
of the spin-triplet Andreev reflection.
We demonstrate a significant increase of the spin-triplet Andreev reflection
when the Ising superconductor is in the single-band case.
In addition, the dependence of the Andreev reflection on the incident energy and incident
angle are investigated. When the incident energy is equal to the superconductor gap,
a complete Andreev reflection can occur
regardless of the Fermi energy (spin polarization) of the ferromagnet.
The spin-triplet Andreev reflection can be strongly enhanced for the suitable oblique incidence.
We also calculate the conductance spectroscopies of both zero bias and finite bias in detail,
and study the influence of gate voltage, exchange energy, and
spin-orbit coupling on the conductance spectroscopy.
A strong magnetoanisotropy with period $\pi$ is shown in the conductance spectroscopy
and a large conductance peak always emerges at the superconductor gap for the case of
the magnetization direction being parallel to the junction plane.
This work offers a comprehensive and systematic study of the spin-triplet
Andreev reflection, and may be helpful for the
detecting of Ising superconductivity in actual experiments, 
and has underlying application of $\pi$-periodic spin valve in spintronics.

\section*{Acknowledgement}
This work was financially supported by National Key R and D Program of China (2017YFA0303301),
NBRP of China (2015CB921102), NSF-China under Grants Nos. 11574007 and 11274364, and
the Key Research Program of the Chinese Academy of Sciences (Grant No. XDPB08-4).

\end{document}